\documentclass[journal]{IEEEtran}

\usepackage{cite}
\usepackage{amsmath,amssymb,amsfonts}
\usepackage{graphicx}
\usepackage{textcomp}

\begin{document}

\title{Analysis of Nonlinear Phase Noise in Coherent Fiber-Optic Systems Based on Phase Shift Keying}

\author{Shiva~Kumar,~\IEEEmembership{Member,~IEEE}%
\thanks{Manuscript received December 16, 2008; revised May 02, 2009. This work was supported by the NSERC discovery Grant. First published July 07, 2009; current version published September 10, 2009.}%
\thanks{The author is with the Department of Electrical and Computer Engineering, McMaster University, Hamilton, ON L8S 4K1 Canada.}}

\markboth{JOURNAL OF LIGHTWAVE TECHNOLOGY, VOL. 27, NO. 21, NOVEMBER 1, 2009}{KUMAR: ANALYSIS OF NONLINEAR PHASE NOISE IN COHERENT FIBER-OPTIC SYSTEMS}

\maketitle

\newcommand\copyrightnotice{%
\begin{bgroup}\small\noindent
\copyright~2009 IEEE. Personal use of this material is permitted. Permission from IEEE must be obtained for all other uses, in any current or future media, including reprinting/republishing this material for advertising or promotional purposes, creating new collective works, for resale or redistribution to servers or lists, or reuse of any copyrighted component of this work in other works. DOI: 10.1109/JLT.2009.2026589
\end{bgroup}\vspace{1em}
}

\begin{abstract}
Analytical expressions for the phase variance in a nonlinear fiber optic system based on phase-shift keying are developed. The Gauss-Hermite functions are used as the orthogonal basis to represent the noise field. Number of degrees of freedom (DOF) to accurately model the phase variance is estimated. The amplifier noise excites higher order Gauss-Hermite noise modes and the nonlinear mixing of a signal pulse and higher order Gauss-Hermite noise mode leads to new noise fields which enhance the nonlinear phase noise. The higher order noise modes propagate linearly and enhance the linear phase noise if the matched filter is not used at the receiver. Analytical expression for the optimum launch power is developed taking into account the linear and nonlinear phase noise.
\end{abstract}


\section{INTRODUCTION}
\IEEEPARstart{T}{he} phase fluctuations induced by the coupling between Kerr nonlinearity of the fiber and amplified spontaneous emission (ASE) of inline amplifiers lead to performance degradation in fiber-optic systems based on phase-shift keying (PSK) or differential phase-shift keying (DPSK) [1]--[4]. The amplitude fluctuations caused by ASE is translated into phase fluctuations because of fiber nonlinearity. This is known as nonlinear phase noise. If the phase deviation exceeds $\pi/2$, it leads to a bit error. This problem is first analyzed by Gordon and Mollenauer [1], and, hence, this noise is also called ``Gordon-Mollenauer phase noise.''

Gordon and Mollenauer pointed out that two degrees of freedom (DOFs) of the noise modes are of importance [1]. These noise modes have the same form as the signal pulse. One of the noise modes is in phase with the signal and the other in quadrature. The in-phase component of the noise changes the amplitude of the signal pulse and, hence, leads to energy change while the quadrature component leads to a linear phase shift. The energy change is translated into an additional phase shift due to fiber nonlinearity. Gordon and Mollenauer argued that the noise modes other than the above mentioned modes have less significant effects if the optical bandwidth is not too large.

Mecozzi analyzed the nonlinear phase noise with arbitrary DOFs [5] and showed that if the signal bandwidth is equal to the noise bandwidth, two DOFs are sufficient to describe the nonlinear phase noise, which corresponds to a matched filter. If the filter bandwidth is larger than signal bandwidth, additional DOFs are required. Later Ho [6], [7] and Mecozzi [8] derived analytical expressions for the probability density function of nonlinear phase noise.

The analytical expressions of Gordon and Mollenauer [1], Mecozzi [5], [8] and Ho [6], [7] are valid only if the fiber dispersion is zero. Attempts have been made to calculate the impact of nonlinear phase noise in the presence of dispersion [9]--[19]. Green et al. [9] showed that variance of nonlinear phase noise becomes quite small in dispersion managed transmission lines when the absolute dispersion of the transmission fiber becomes large. Green et al. assumed that the signal is CW and they used the approach typically used in the study of modulational instability. Later in [10], the variance of nonlinear phase noise is calculated for a Gaussian pulse in a dispersion managed transmission line and results showed that variance of nonlinear phase noise due to single pulse self phase modulation (SPM) is quite small as compared to the case of no dispersion. McKinstrie et al. [11], [12] and Hanna et al. [13] developed analytical expressions for the nonlinear phase variance in soliton-based systems. Ho and Wang [14], [15] used first order perturbation technique to calculate the nonlinear phase noise variance due to self-phase modulation (SPM) as well as intra-channel cross-phase modulation (IXPM). Zhang et al. [16] used variational approach to calculate the phase variance in DPSK systems and showed that the IXPM causes a partial correlation between adjacent pulses' phase noise, leading to a decrease instead of an increase in differential phase noise as compared to the case of SPM acting alone. Serena et al. [17] used a parametric gain approach to study the nonlinear phase noise and they calculated the bit error rate (BER) using Karhunen-Loeve method. Their results showed that the 3-dB advantage of DPSK systems over OOK systems, based on linear considerations, is lost for large nonlinear phases due to the larger sensitivity of DPSK to nonlinear phase noise. Demir [19] used a linearized perturbation theory to characterize nonlinear phase noise and found that the ASE noise which is initially white becomes colored because of severe disastrous amplification of ASE noise by the Kerr nonlinear effects. An extensive review of the previous work on nonlinear phase noise can also be found in [19].

In this paper, we use a technique based on Gauss-Hermite basis functions to calculate the variance of phase noise in a coherent system based on phase shift keying (PSK). In [10], the variance of nonlinear phase noise of a dispersive nonlinear fiber-optic system is calculated including only two degrees of freedoms (DOFs). Here, we extend the analysis to include arbitrary DOFs. DOF is the number of orthonormal set of functions to describe the noise field. In [1] and [10], two DOFs that have the same form as the signal pulse are considered. In a linear system, when a matched filter is used at the end of the transmission line, these two DOFs are sufficient to describe the total noise field since the other noise modes are orthogonal to the signal and do not contribute. In [1] and [10], it is assumed that other DOFs are less significant when a matched filter is used even for the nonlinear systems. In this paper, we investigate the impact of the other DOFs and find that the maximum error introduced by including only two DOFs is about 10\% in our parameter space when a matched filter is used. However, this error could increase for longer transmission distance/higher launch power.

We represent the noise field as the superposition of Gauss-Hermite functions. Gauss-Hermite functions are the exact modes of the paraxial Helmholtz equation and they are widely used in the description of optical resonators [20], [21]. Lazaridis et al. [22] showed that the Gauss-Hermite functions are the exact solutions of the linear part of the nonlinear Schr\"{o}dinger (NLS) equation used to describe the temporal pulse propagation. Turitsyn and Mezentsev [23] used chirped Gauss-Hermite orthogonal functions for the description of the breathing dynamics of dispersion-managed solitons. Lakoba and Kaup [24] represented a pulse in the strongly dispersion managed fiber as a linear superposition of Gauss-Hermite functions and they obtained the conditions for stationary nonlinear pulse propagation. If a $j$th order Gauss-Hermite mode is excited at the fiber input, its pulse shape remains invariant with distance (assuming nonlinear effects are absent), but the pulse width and chirp vary because of fiber dispersion. Therefore, Gauss-Hermite functions provide a convenient set of orthogonal basis to represent both signal and noise fields at any transmission distance. We assume that the signal launched to the fiber is a Gaussian pulse. The complex amplitude of the zeroth order Gauss-Hermite noise mode (which is also a Gaussian pulse) modifies the signal amplitude and is responsible for linear and nonlinear phase noise. The nonlinear mixing of the higher order noise modes with the signal pulse leads to new noise fields. Although the higher order noise modes are orthogonal to signal when matched filters are used, the noise fields generated by the parametric process are not. In fact, some of them occupy the same spectral region as the signal pulse leading to nonlinear amplitude and phase noise. As the signal power and/or transmission distance increases, the amplitudes of the noise fields generated by the parametric process increases, too. We have also investigated the impact of electrical Gaussian filters of arbitrary bandwidths at the end of the transmission line. As the filter bandwidth exceeds the signal bandwidth, higher order noise modes would not be orthogonal to the signal and they contribute to the linear as well as nonlinear phase noise.

Gordon and Mollenauer showed that the total phase variance can be minimized if the mean nonlinear phase shift is approximately 1 rad. We have found that it holds true even for the dispersive nonlinear systems, if the mean nonlinear phase of the signal sample after the correlator is approximately 1 rad. However, our expression for the mean nonlinear phase takes into account the dispersive effects. Based on this, an analytic expression for the optimum launch power is developed assuming two DOFs and matched filters.

In this paper, we make the following assumptions/approximations so that the analytical expressions become simple, and scaling laws are clear. (i) The transmission system is operating in the pseudo-linear regime and the nonlinear effects can be treated as a first order perturbation on the linear system. (ii) The signal power is much larger than the mean ASE power added by amplifiers. This approximation could break down if the launch power is too low and/or amplifier noise figure and gain are quite large. (iii) Interaction between dispersion, SPM and ASE on a single pulse is considered. The interaction between IXPM and ASE, although it could be important in some system set up, is not analyzed here.

The paper is organized as follows. In Section II, analytical description of nonlinear phase noise is provided. A simple analytical expression for the phase variance is derived for arbitrary dispersion maps when the DOF is two. Including the noise fields generated by the mixing of signal and higher order noise modes, analytical expressions are obtained for the case of arbitrary DOFs. In Section III, the Monte-Carlo simulations of the NLS equation is carried out and the analytical results are cross-validated. Section IV contains the summary of this work.

\section{NONLINEAR PHASE NOISE}
The optical field envelope in an amplified transmission system is governed by the nonlinear Schr\"{o}dinger (NLS) equation in the lossless form
\begin{equation}
i\frac{\partial u}{\partial z}-\frac{\beta_{2}(z)}{2}\frac{\partial^{2}u}{\partial t^{2}}=-\gamma \exp[-w(z)]|u|^{2}u+iR(z,t)
\end{equation}
where $\beta_{2}(z)$ is the dispersion profile, $\gamma$ is the nonlinear coefficient, $w(z)=\int_{0}^{z}\alpha(s)ds$, $\alpha(z)$ is the fiber loss/amplifier gain profile. We assume that the fiber loss is exactly compensated by an inline amplifier. $R$ represents the noise field due to amplification, i.e.,
\begin{equation}
R(z,t)=\sum_{m=1}^{N_{a}}\delta(z-L_{m})n^{(m)}(t)
\end{equation}
where $L_{m}$ is the location of an amplifier, $N_{a}$ is the number of amplifiers, and $n^{(m)}(t)$ is the noise field due to an amplifier located at $L_{m}$. The mean and autocorrelation function of the noise field are given by
\begin{align}
\langle n^{(m)}(t)\rangle &= 0, \\
\langle n^{(m)}(t)n^{(m)*}(t')\rangle &= \rho_{m}\delta(t-t') \\
\langle n^{(m)}(t)n^{(m)}(t')\rangle &= 0
\end{align}
where $\rho_{m}$ is the Amplified Spontaneous Emission (ASE) power spectral density per polarization of an amplifier located at $L_{m}$ given by [1]
\begin{equation}
\rho_{m}=n_{sp}h\bar{\nu}(G_{m}-1)
\end{equation}
where $G_{m}$ is the gain of the amplifier, $n_{sp}$ is spontaneous noise factor, $h$ is Planck's constant and $\bar{\nu}$ is the mean optical carrier frequency. In the absence of nonlinear effects and amplifier noise, if a Gaussian pulse is launched to the fiber, its propagation is given by [25]
\begin{equation}
u_{lin}(z,t)=\sqrt{E}F(z,t)
\end{equation}
\begin{equation}
F(z,t)=\left[\frac{p(z)}{\sqrt{\pi}}\right]^{1/2}\exp\left\{-\frac{[p^{2}(z)+iC(z)]t^{2}}{2}+i\theta_{0}(z)\right\}
\end{equation}
where $E$ is the pulse energy, $p(z)$, $C(z)$ and $\theta_{0}(z)$ are the inverse pulse width, chirp and phase factor, respectively, given by
\begin{align}
p(z) &= \frac{T_{0}}{\sqrt{T_{0}^{4}+S^{2}(z)}}, \quad C(z) = \frac{S(z)p^{2}(z)}{T_{0}^{2}}, \\
\theta_{0}(z) &= \frac{1}{2}\tan^{-1}\left[\frac{S(z)}{T_{0}^{2}}\right].
\end{align}
Here, $T_{0}$ is the half-width at $1/e$-intensity point, and $S(z)$ is the accumulated dispersion
\begin{equation}
S(z)=\int_{0}^{z}\beta_{2}(s)ds.
\end{equation}
The peak power, $P$ and energy, $E$ are related by
\begin{equation}
P=\frac{E}{T_{eff}}
\end{equation}
where $T_{eff}=\sqrt{\pi}T_{0}$ and $F(z,t)$ is normalized such that $\int_{-\infty}^{\infty}|F(z,t)|^{2}dt=1$.

Expanding the optical field in a series, we have
\begin{equation}
u(z,t)=u^{(0)}(z,t)+\gamma u^{(1)}(z,t)+\gamma^{2}u^{(2)}(z,t)+\dots
\end{equation}
where $u^{(j)}(z,t)$, $j\neq0$ is the $j$th order correction due to fiber nonlinearity, and $u^{(0)}(z,t)$ is the zeroth order linear solution. In this paper, we focus only up to the first order correction to the optical field envelope. Substituting (13) in (1) (without the amplifier noise term) and collecting the terms proportional to $\gamma$, we obtain
\begin{equation}
i\frac{\partial u^{(1)}}{\partial z}-\frac{\beta_{2}(z)}{2}\frac{\partial^{2}u^{(1)}}{\partial t^{2}}=-\exp[-w(z)]|u^{(0)}|^{2}u^{(0)}.
\end{equation}
We will use (14) in Section II-A and B to calculate the impact of SPM on the signal and noise fields.

\subsection{Two DOFs}
Consider the optical field envelope immediately after an amplifier located at $L_{m}$. Focusing only on the impact of the noise added by this amplifier, the linear part of the optical field envelope at $z=L_{m+}$ can be written as
\begin{equation}
u_{lin}(L_{m+},t)=u_{lin}(L_{m},t)+n(t)
\end{equation}
where $n(t)\equiv n^{(m)}(t)$ is the noise field added by the amplifier at $L_{m}$. In this subsection, we consider only two noise modes given by
\begin{equation}
n(t)=[n_{0r}+in_{0i}]F(L_{m},t)
\end{equation}
where $n_{0r}=\text{Re}[n_{0}]$ and $n_{0i}=\text{Im}[n_{0}]$ are the amplitudes of the in-phase component and quadrature components, respectively, and $n_{0}$ is the complex amplitude of the noise field. In (16), the in-phase and quadrature components are assumed to have the same form as the signal pulse. Substituting (16) in (15), we find
\begin{equation}
u^{(0)}(L_{m+},t)=(\sqrt{E}+n_{0})F(L_{m},t).
\end{equation}
Thus, the complex amplitude of the field envelope has changed because of the amplifier noise. Treating (17) as the initial condition, the zeroth order optical field envelope is described by
\begin{equation}
u^{(0)}(z,t)=(\sqrt{E}+n_{0})F(z,t), \quad z>L_{m}.
\end{equation}
Substituting (18) in (14), the first-order correction due to SPM can be written as
\begin{multline}
i\frac{\partial u^{(1)}}{\partial z}-\frac{\beta_{2}(z)}{2}\frac{\partial^{2}u^{(1)}}{\partial t^{2}}=-\exp[-w(z)] \\
\times|(\sqrt{E}+n_{0})F(z,t)|^{2}(\sqrt{E}+n_{0})F(z,t).
\end{multline}
In practical systems operating in the pseudo-linear regime, the dispersion of the transmission fibers is fully compensated at the receiver either in optical or electrical domain, i.e., $S(L_{tot})=0,$ where $L_{tot}$ is the total transmission distance. Solving (19) with the condition, $S(L_{tot})=0$ we find [26]--[28]
\begin{equation}
u^{(1)}(L_{tot},t)=i(\sqrt{E}+n_{0})F(0,t)(E+\delta E)g(L_{m},t)
\end{equation}
where
\begin{align}
\delta E &= 2\sqrt{E}n_{0r} \\
g(z,t) &= \frac{T_{0}}{\sqrt{\pi}}\int_{z}^{L_{tot}}\frac{\exp[-w(r)-\Delta(r)t^{2}]dr}{\sqrt{T_{0}^{4}+3S^{2}(r)+2iT_{0}^{2}S(r)}} \\
\Delta(r) &= \frac{T_{0}^{2}-iS(r)}{T_{0}^{2}[T_{0}^{2}+i3S(r)]}.
\end{align}
In (20), we have ignored the higher order terms such as $n_{0r}^{2}$, and $n_{0i}^{2}$ under the assumption that the noise power is much smaller than the signal power. Combining the first order and zeroth order solutions [(18) and (20)], total field envelope at the end of the transmission line is
\begin{equation}
u(L_{tot},t)=(\sqrt{E}+n_{0})F(0,t)[1+i\gamma(E+\delta E)g(L_{m},t)].
\end{equation}
From (21) and (24), we see that the in-phase noise component $n_{0r}$ is responsible for energy shift and the consequent nonlinear phase shift.

The output of in-phase-quadrature (IQ) coherent receiver has two output currents which are proportional to real and imaginary parts of the optical field envelope. For simplicity, we assume that the constant of proportionality to be unity since it does not affect the performance of a long-haul fiber-optic systems. The output of the IQ receiver passes through the digital signal processing (DSP) unit after A/D converter and arbitrary filter shapes can be realized using DSP. In this section, we assume that the output of the IQ receiver passes through a matched filter, or equivalently a correlator and the decision is based on the correlator output
\begin{align}
u_{f} &= \int_{-T_{b}/2}^{T_{b}/2}u(L_{tot},t)F^{*}(0,t)dt \nonumber \\
&\approx \int_{-\infty}^{\infty}u(L_{tot},t)F^{*}(0,t)dt
\end{align}
where $T_{b}$ is the bit interval. Substituting (8) and (24) in (25), we find
\begin{equation}
u_{f}=(\sqrt{E}+n_{0})[1+i\gamma(E+\delta E)g_{f}(L_{m})]
\end{equation}
where
\begin{align}
g_{f}(L_{m}) &= \frac{T_{0}}{\sqrt{\pi}}\int_{L_{m}}^{L_{tot}}G(r)dr \\
G(r) &= \frac{\exp[-w(r)]}{\sqrt{[1+T_{0}^{2}\Delta(r)][T_{0}^{4}+3S^{2}(r)+2iT_{0}^{2}S(r)]}}.
\end{align}
The phase of the correlator output is
\begin{align}
\phi &= \tan^{-1}\frac{\text{Im}[u_{f}]}{\text{Re}[u_{f}]} \nonumber \\
&\approx \gamma Eg_{fr}(L_{m})+\gamma\delta Eg_{fr}(L_{m})+\frac{n_{0i}}{\sqrt{E}}
\end{align}
where $g_{fr}(L_{m})=\text{Re}[g_{f}(L_{m})]$. In (29), we have ignored terms proportional to $\gamma^{2}$, $n_{0r}^{2},$ $n_{0i}^{2}$ and $n_{0r}n_{0i}$. The first, second and the last terms on the right hand side of (29) represent the deterministic nonlinear phase change, nonlinear and the linear phase changes due to ASE of the amplifier located at $L_{m}$, respectively. Therefore, the phase changes due to ASE of the amplifier located at $L_{m}$ are
\begin{equation}
\delta\phi_{m}=\gamma\delta Eg_{fr}(L_{m})+\frac{n_{0i}}{\sqrt{E}}.
\end{equation}
Variance of energy shift is related to the variance of $n_{0r}$. From (4), (5) and (21), we have
\begin{align}
\langle n_{0r}^{2}\rangle &= \langle n_{0i}^{2}\rangle = \frac{\rho_{m}}{2} \\
\langle\delta E^{2}\rangle &= 2\rho_{m}E
\end{align}
Equation (32) holds because Ito calculus applies for the signal and noise set up considered here. Squaring and averaging (30), and using (31)--(32), we obtain
\begin{equation}
\langle\delta\phi_{m}^{2}\rangle=2\rho_{m}E[\gamma g_{fr}(L_{m})]^{2}+\frac{\rho_{m}}{2E}.
\end{equation}
The first and the second terms in (33) represent the variance of nonlinear phase noise and linear phase noise, respectively, due to the amplifier located at $L_{m}$. The in-phase component of the noise field, $n_{0r}$ and the quadrature component, $n_{0i}$ are responsible for the nonlinear and linear phase noise, respectively. Since these components are statistically independent, total variance is the sum of the variance due to linear and nonlinear phase noise, as given by (33).

Since the noise of amplifiers is statistically independent, variance of phase noise due to all the amplifiers is
\begin{equation}
\langle\delta\phi^{2}\rangle=\sum_{m=1}^{N_{a}}\langle\delta\phi_{m}^{2}\rangle.
\end{equation}

For M-ary PSK signals, symbol error probability is determined solely by the probability density function (PDF) of the phase. Under the Gaussian PDF assumption, the variance calculated from (34) can be related to the symbol error probability for MPSK signals [29]. Equation (34) is valid for arbitrary dispersion maps with $S(L_{tot})=0.$ To simplify (34) further and also to make a direct comparison with [1] and [10], we consider a transmission fiber consisting of two segments of equal lengths within an amplifier spacing. The dispersion of the first segment is anomalous whereas that of the second segment is equal in magnitude but opposite in sign. We assume that there is no pre- and postcompensation of dispersion. Since the amplifier spans are identical, $L_{m}=mL$, $m=1,2,\dots,N_{a}$, where $L$ is the amplifier spacing, we can write
\begin{align}
g_{f}(L_{m}) &= (N_{a}-m)h_{f} \\
h_{f} &= \frac{T_{0}}{\sqrt{\pi}}\int_{0}^{L}G(r)dr
\end{align}
and (33) is modified as
\begin{equation}
\langle\delta\phi_{m}^{2}\rangle=2\rho E[\gamma(N_{a}-m)h_{fr}]^{2}+\frac{\rho}{2E}
\end{equation}
where $h_{fr}=\text{Re}[h_{f}]$ and $\rho_{m}=\rho.$ The first term in (37) is same as the expression for the nonlinear phase variance derived in [10] except that the impact of matched filter on the first order correction was not included in [10] and terms second order in $\gamma$ are ignored here. Adding contributions to the phase variance from all the amplifiers, we obtain the total variance as
\begin{equation}
\langle\delta\phi^{2}\rangle=\frac{N_{a}(N_{a}-1)(2N_{a}-1)\rho E(\gamma h_{fr})^{2}}{3}+\frac{\rho N_{a}}{2E}.
\end{equation}
When $N_{a}\gg1$, (38) can be approximated as
\begin{equation}
\langle\delta\phi^{2}\rangle\approx\frac{2\rho E(\gamma h_{fr})^{2}N_{a}^{3}}{3}+\frac{\rho N_{a}}{2E}.
\end{equation}

The variance of nonlinear phase noise grows cubically with number of amplifiers while the variance of linear phase noise increases linearly for large $N_{a}$. Therefore, nonlinear phase noise could become dominant for ultra long haul transmission systems. In Section III, we will use (38) and (34) to calculate the variance of phase noise.

The optimum launch power is calculated by differentiating $\langle\delta\phi^{2}\rangle$ with respect to $E$ and setting it to zero. Since the peak power, $P=E/T_{eff}$, we find the optimum launch power as
\begin{equation}
P_{opt}=\frac{1}{\gamma h_{fr}T_{eff}}\sqrt{\frac{3}{2(N_{a}-1)(2N_{a}-1)}}.
\end{equation}

In a dispersion-free system, Gordon and Mollenauer showed that the total phase variance is minimized when the mean nonlinear phase shift is approximately 1 rad. It holds true even for the dispersive systems, if the mean nonlinear phase of the signal sample used for decision, $\gamma h_{fr}N_{a}E$ is approximately 1 rad.

\subsection{Arbitrary DOFs}
In this section, we consider the impact of an electrical Gaussian filter of arbitrary bandwidth. An exact solution of the linear part of NLS equation ($\gamma=0$ in (1)) is described by Gauss-Hermite function of the form
\begin{align}
H_{j}(t) &= k_{j}\exp\left(\frac{t^{2}}{2}\right)\frac{d^{j}}{dt^{j}}\exp(-t^{2}) \\
k_{j} &= \frac{(-1)^{j}}{\sqrt{2^{j}j!\sqrt{\pi}}}.
\end{align}
If the initial optical field envelope at $z=0$ is $H_{j}(t/T_{0})/\sqrt{T_{0}}$, the field envelope at any $z$ is given by [22] (Appendix A)
\begin{equation}
\psi_{j}(z,t)=\sqrt{p(z)}H_{j}(pt)\exp\left[iC(z)\frac{t^{2}}{2}+i(2j+1)\theta_{0}(z)\right]
\end{equation}
where the evolution of $p(z)$, $C(z)$ and $\theta_{0}(z)$ are given by (9) and (10). In other words, the evolution of pulse width and chirp of a $j$th order Gauss-Hermite mode is same as that of a Gaussian pulse, $F(z,t)$ (which is a zeroth order Gauss-Hermite mode) and the phase factor is $(2j+1)$ times $\theta_{0}(z)$. Gauss-Hermite functions form an orthonormal set and arbitrary linear signal field, $s(z,t)$ can be expressed as the superposition of $\psi_{j}(z,t)$
\begin{equation}
s(z,t)=\sum_{j=0}^{\infty}s_{j}\psi_{j}(z,t)
\end{equation}
where $s_{j}$ is the amplitude of the $j$th order signal mode, given by
\begin{equation}
s_{j}=\int_{-\infty}^{\infty}s(z,t)\psi_{j}^{*}(z,t)dt.
\end{equation}
We assume that the pulse launched to the fiber is a Gaussian pulse of the form given by (8). In this case, $s_{0}=\sqrt{E}$ and $s_{j}=0, j>0$. Next, as in the previous section, we consider the noise field added by the amplifier located at $L_{m}$. The noise field can also be written as a superposition of $\psi_{j}(z,t)$
\begin{equation}
n(z,t)=\sum_{j=0}^{\infty}n_{j}\psi_{j}(z,t), \quad z \ge L_{m}
\end{equation}
where $n_{j}$ is the amplitude of the $j$th order noise mode given by
\begin{equation}
n_{j}=\int_{-\infty}^{\infty}n(L_{m},t)\psi_{j}^{*}(L_{m},t)dt, \quad z \ge L_{m}
\end{equation}
and $n(L_{m},t)\equiv n^{(m)}(t)$. If we take only the first term ($j=0$) in the summation on the right hand side of (46), it corresponds to two DOFs ($n_{0r}$ and $n_{0i}$) discussed in the previous subsection. We ignore the terms of the summation in (46) after $j>J$ which corresponds to $2(J+1)$ DOFs. Multiplying $n_{j}$ and $n_{k}$ and using (3)--(5) and (47), it follows that
\begin{align}
\langle n_{j}\rangle &= 0 \\
\langle n_{j}n_{k}^{*}\rangle &= \rho_{m}\delta_{jk}, \quad \langle n_{j}n_{k}\rangle = 0
\end{align}
where $\delta_{jk}$ is a Kronecker delta function.

The linear part of the optical field envelope for $z \ge L_{m}$ is
\begin{align}
u^{(0)}(z,t) &= s(z,t)+n(z,t) \nonumber \\
&= \sum_{j=0}^{J}b_{j}\psi_{j}(z,t)
\end{align}
where
\begin{align}
b_{0} &= \sqrt{E}+n_{0} \\
b_{j} &= n_{j}, \quad j>0.
\end{align}

Nonlinear mixing of signal and noise modes leads to new noise fields which can be calculated using the first order perturbation theory. Choosing the linear part of the optical field envelope given by (50) as the zeroth order solution, the first order solution can be found out using (14). Using (50), the term on the right hand side of (14) can be written as
\begin{equation}
|u^{(0)}|^{2}u^{(0)}=\sum_{j=0}^{J}\sum_{k=0}^{J}\sum_{l=0}^{J}b_{j}b_{k}b_{l}^{*}\psi_{j}\psi_{k}\psi_{l}^{*}
\end{equation}
Now we assume that the signal power is much larger than the mean noise power, i.e.,
\begin{equation}
\sqrt{E} \gg \langle|n_{j}|\rangle, \quad j=0,1,\dots,J.
\end{equation}
From (54), it follows that the terms proportional to $\sqrt{E}\langle|n_{j}|^{2}\rangle$ are much smaller than the terms proportional to $E\langle|n_{j}|\rangle.$ In other words, only the terms proportional to $E\sqrt{E}$ and $E[n_{k}]$ in (53) contribute significantly (after averaging) and the other terms can be ignored. Now, (14) can be written as
\begin{multline}
i\frac{\partial u^{(1)}}{\partial z}-\frac{\beta_{2}(z)}{2}\frac{\partial^{2}u^{(1)}}{\partial t^{2}}=-E\exp[-w(z)] \\
\times\left[(\sqrt{E}+2n_{0}+n_{0}^{*})|\psi_{0}|^{2}\psi_{0}+\sum_{j=1}^{J}\left(2n_{j}|\psi_{0}|^{2}\psi_{j}+n_{j}^{*}\psi_{0}^{2}\psi_{j}^{*}\right)\right].
\end{multline}
Here, $u^{(1)}$ represents the new noise fields generated by the mixing of signal and noise fields. Since (55) is a linear equation, total solution, $u^{(1)}(z,t)$ can be expressed as the superposition of responses due to different forcing functions on the right hand side of (55), i.e.,
\begin{equation}
u^{(1)}(z,t)=\sum_{j=0}^{J}\psi_{j}^{(1)}(z,t)
\end{equation}
where $\psi_{j}^{(1)}(z,t)$ is the noise field generated by the nonlinear mixing of the signal (with energy $E$) and the $j$th order noise mode. Substituting (56) in (55), we obtain
\begin{multline}
i\frac{\partial\psi_{0}^{(1)}}{\partial z}-\frac{\beta_{2}(z)}{2}\frac{\partial^{2}\psi_{0}^{(1)}}{\partial t^{2}}=-E\exp[-w(z)] \\
\times(\sqrt{E}+2n_{0}+n_{0}^{*})|\psi_{0}|^{2}\psi_{0}
\end{multline}
and
\begin{multline}
i\frac{\partial\psi_{j}^{(1)}}{\partial z}-\frac{\beta_{2}(z)}{2}\frac{\partial^{2}\psi_{j}^{(1)}}{\partial t^{2}}=-E\exp[-w(z)] \\
\times\left[2n_{j}|\psi_{0}|^{2}\psi_{j}+n_{j}^{*}\psi_{0}^{2}\psi_{j}^{*}\right], \quad j>0.
\end{multline}

The solution of (57) is given by (20) of Section II-A. Equation (58) is solved in Appendix B. In this subsection, we assume that the filter bandwidth is arbitrary. For the sake of computational ease, we choose the electrical Gaussian filter at the receiver given by
\begin{equation}
H_{G}(f)=\exp\left(\frac{-f^{2}}{2f_{0}^{2}}\right)
\end{equation}
When $f_{0}=1/(2\pi T_{0})$, the sample of the filter output at $t=0$ corresponds to the output of the correlator discussed in the Section II-A. Solving (58) and after passing it through the filter, the sample of the nonlinearly generated noise fields at $t=0$ is (Appendix B)
\begin{equation}
\psi_{jf}^{(1)}=E(n_{j}A_{j}+n_{j}^{*}B_{j}), \quad j>0
\end{equation}
where $A_{j}$ and $B_{j}$ are defined in Appendix B. From (107), (113), (112), (120), and (119), we find that $A_{j}$ and $B_{j}$ are zero when $j$ is odd which means that the energy transfer from the initially launched fundamental signal mode (Gaussian) to a higher order odd noise mode does not occur within the first order approximations.

The higher order noise modes given by the terms with $j>0$ in (50) also pass through the filter, and, therefore, the corresponding filter output at $t=0$ is
\begin{equation}
n_{f}=\sum_{j=1}^{J}n_{j}Z_{j}
\end{equation}
where
\begin{equation}
Z_{j}=\int_{-\infty}^{\infty}\mathcal{F}[\psi_{j}(L_{tot},t)]\exp\left[-\frac{f^{2}}{2f_{0}^{2}}\right]df, \quad j=0,1,2,\dots
\end{equation}
where $\mathcal{F}$ denotes the Fourier transform. Substituting (41) in (62), $Z_{j}$ is calculated as
\begin{align}
Z_{j} &= \frac{2\pi(-i)^{j}a^{j/2}f_{0}T_{0}(j-1)!}{\sqrt{j!}(1+a)^{(j+1)/2}(\frac{j}{2}-1)!2^{j/2-1}}, \quad \text{when } j \text{ is even} \nonumber \\
&= 0, \quad \text{when } j \text{ is odd}
\end{align}
where
\begin{equation}
a=\frac{[4\pi^{2}(f_{0}T_{0})^{2}-1]}{2}.
\end{equation}
When a matched filter is used, $a=0$, $Z_{0}=1$ and $Z_{j}=0$ for $j>0$.

The zeroth order solution at $t=0$ after passing through the filter is
\begin{equation}
u_{f}^{(0)}=(\sqrt{E}+n_{0})Z_{0}+n_{f}.
\end{equation}

When $j=0$, the first order correction is given by (20) and after passing through the Gaussian filter, the sample at $t=0$ is given by
\begin{equation}
\psi_{0f}^{(1)}=i(\sqrt{E}+n_{0})(E+\delta E)g_{f}(L_{m})
\end{equation}
where
\begin{equation}
g_{f}(L_{m})=\int_{-\infty}^{\infty}\mathcal{F}[g(L_{m},t)]\exp\left[-\frac{f^{2}}{2f_{0}^{2}}\right]df.
\end{equation}
When a matched filter is used ($a=0$), (67) and (27) are identical. Combining (60), (61), (65), and (66), total filter output at $t=0$ is
\begin{align}
u_{f} &= u_{f}^{(0)}+\gamma\sum_{j=0}^{J}\psi_{jf}^{(1)} \nonumber \\
&= (\sqrt{E}+n_{0})[Z_{0}+i\gamma(E+\delta E)g_{f}(L_{m})] \nonumber \\
&\quad + \sum_{j=1}^{J}\left[n_{j}(Z_{j}+\gamma EA_{j})+n_{j}^{*}\gamma EB_{j}\right].
\end{align}

The output phase is
\begin{align}
\phi &= \tan^{-1}\frac{\text{Im}[u_{f}]}{\text{Re}[u_{f}]} \nonumber \\
&\approx \frac{\gamma(E+\delta E)g_{fr}(L_{m})}{Z_{0}}\left[1-\sum_{j=1}^{J}\frac{n_{jr}Z_{j}}{\sqrt{E}Z_{0}}\right]+\frac{n_{0i}}{\sqrt{E}} \nonumber \\
&\quad + \sum_{j=1}^{J}\frac{n_{ji}Z_{j}}{\sqrt{E}Z_{0}}\left[1+\frac{\gamma Eg_{fi}(L_{m})}{Z_{0}}\right] \nonumber \\
&\quad + \sum_{j=1}^{J}\frac{\gamma\sqrt{E}}{Z_{0}}[n_{jr}P_{j}+n_{ji}Q_{j}]
\end{align}
where
\begin{align}
P_{j} &= \text{Im}[A_{j}+B_{j}] \\
Q_{j} &= \text{Re}[A_{j}-B_{j}]
\end{align}
$n_{jr}=\text{Re}(n_{j})$, $n_{ji}=\text{Im}(n_{j})$, and $g_{fi}=\text{Im}(g_{f})$. In (69), the first term ($\propto E$) represents the deterministic nonlinear phase change due to SPM. The phase changes due to ASE introduced by the $m$th amplifier can be written as
\begin{equation}
\delta\phi_{m}=\delta\phi_{Lin}+\delta\phi_{NL}
\end{equation}
where $\delta\phi_{Lin}$ and $\delta\phi_{NL}$ are the linear and nonlinear phase changes, respectively, and are given by
\begin{align}
\delta\phi_{Lin} &= \frac{1}{\sqrt{E}Z_{0}}\sum_{j=0}^{J}n_{ji}Z_{j} \\
\delta\phi_{NL} &= \frac{\gamma\sqrt{E}}{Z_{0}}\left[2n_{0r}g_{fr}(L_{m})+\sum_{j=1}^{J}(n_{jr}P_{j}'+n_{ji}Q_{j}')\right]
\end{align}
where
\begin{align}
P_{j}'(L_{m}) &= P_{j}(L_{m})-\frac{g_{fr}(L_{m})Z_{j}}{Z_{0}} \\
Q_{j}'(L_{m}) &= Q_{j}(L_{m})+\frac{g_{fi}(L_{m})Z_{j}}{Z_{0}}.
\end{align}

From (74), we see that contributions to the nonlinear phase noise comes not only from the real part of $n_{j}$ but also from imaginary part of $n_{j}$ whereas in the case of two DOFs (Section II-A), only the real part of $n_{0}$ is responsible for nonlinear phase noise. The contribution to the linear phase noise comes only from the imaginary part of $n_{j}$ in both cases. Therefore, the linear and nonlinear phase noises are not statistically independent unlike the case in the Section II-A. Squaring and averaging (72), and using (49), we obtain
\begin{multline}
\langle\delta\phi_{m}^{2}\rangle=\frac{\rho_{m}E\gamma^{2}}{Z_{0}^{2}}\left[2g_{fr}^{2}(L_{m})+\sum_{j=1}^{J}\frac{P_{j}'^{2}+Q_{j}'^{2}}{2}\right] \\
+\frac{\rho_{m}}{2E}\left[1+\sum_{j=1}^{J}\frac{Z_{j}^{2}}{Z_{0}^{2}}\right]+\sum_{j=1}^{J}\frac{\gamma\rho_{m}Q_{j}'Z_{j}}{Z_{0}^{2}}.
\end{multline}

The first term ($\propto\gamma^{2}$) on the right hand side of (77) represents the nonlinear phase noise, the second term represents linear phase noise, and the last term represents the correlation between linear and nonlinear phase noise which is absent when $\text{DOF}=2$. The variance of phase noise due to all the amplifier is given by (34).

As before, we consider the simple periodic dispersion map with two types of fiber segments and the average dispersion between amplifiers is zero. In this case, we have
\begin{align}
A_{j}(L_{m}) &= (N_{a}-m)X_{j} \\
B_{j}(L_{m}) &= (N_{a}-m)Y_{j} \\
X_{j}(z) &= \int_{0}^{L}\eta_{j}(z)dz \\
Y_{j}(z) &= \int_{0}^{L}\xi_{j}(z)dz
\end{align}
and $\eta_{j}(z)$ and $\xi_{j}(z)$ are defined in Appendix B. Using (77), (34), (78), and (79), the total phase variance is obtained as
\begin{multline}
\langle\delta\phi^{2}\rangle=\frac{\mu(N_{a})\rho E\gamma^{2}}{Z_{0}^{2}}\left[h_{fr}^{2}+\sum_{j=1}^{J}\frac{U_{j}^{2}+V_{j}^{2}}{4}\right] \\
+\frac{\rho N_{a}}{2E}\left(1+\sum_{j=1}^{J}\frac{Z_{j}^{2}}{Z_{0}^{2}}\right)+\frac{\gamma N_{a}(N_{a}-1)}{2Z_{0}^{2}}\sum_{j=1}^{J}Z_{j}V_{j}
\end{multline}
where
\begin{align}
\mu(N_{a}) &= \frac{N_{a}(N_{a}-1)(2N_{a}-1)}{3} \\
U_{j} &= \text{Im}(X_{j}+Y_{j})-\text{Re}(h_{f})\frac{Z_{j}}{Z_{0}} \\
V_{j} &= \text{Re}(X_{j}-Y_{j})+\text{Im}(h_{f})\frac{Z_{j}}{Z_{0}}.
\end{align}

Comparing the variance of linear phase noise for the case of two DOFs (second term of (38)) and that for the case of $2(J+1)$ DOFs (second term of (82)), we see that the variance is enhanced due to the excitation of higher order noise modes. However, when a matched filter is used, $Z_{j}=0$ for $j>0$ and these two expressions for the linear phase variance become identical. Hence, for the linear case, it is sufficient to include two DOFs when a matched filter is used. However, nonlinear phase noise due to the mixing of signal and higher order noise modes (described by terms $U_{j}$ and $V_{j}$) does not vanish even when the matched filter is used. Equation (82) is used in Section III to calculate the phase variance.

\section{RESULTS AND DISCUSSION}
To test the validity of the approximations done in obtaining (38) and (82), numerical simulations of the nonlinear Schr\"{o}dinger (NLS) equation by the split-step Fourier technique is carried out. We assume the following parameters throughout the paper: nonlinear coefficient $=2.43\text{ W}^{-1}\text{km}^{-1}$, fiber loss coefficient $=0.2\text{ dB/km}$, bit rate $=40\text{ Gb/s}$, $n_{sp}=1$ which corresponds to a noise figure of 3 dB, and spacing between inline amplifiers $=80\text{ km}$. We assume that a Gaussian pulse with full width half-maximum (FWHM) of 12.5 ps is launched to the fiber link so that $T_{0}=7.5\text{ ps}$. The computational bandwidth is 320 GHz and ASE is propagated over the entire computational bandwidth. A Gaussian filter of arbitrary bandwidth is used in electrical domain and no optical filter is used. 4000 runs of NLS equation are carried out and the phase variance of the decision variable is calculated. In Fig. 1, the matched filter is used at the end of the transmission line with $f_{0}=1/(2\pi T_{0})$. For Figs. 1--4, two types of fibers are used between inline amplifiers, the first one is an anomalous dispersion fiber of length 40 km and the second one is the normal dispersion fiber of the same absolute dispersion and the same length. The `+' marks in Fig. 1 shows the numerical simulation results and the solid line shows the analytical results calculated using (82) with $J=6$. As the dispersion increases, the variance of nonlinear phase noise due to SPM decreases consistent with the results of [9] and [10]. The nonlinear phase variance grows cubically with distance and, therefore, the difference between the variances for the case of $|D|=4\text{ ps/nm}\cdot\text{km}$ and $|D|=10\text{ ps/nm}\cdot\text{km}$ increases significantly for longer transmission lengths.

\begin{figure}[htbp]
\centering
\includegraphics[width=0.85\linewidth]{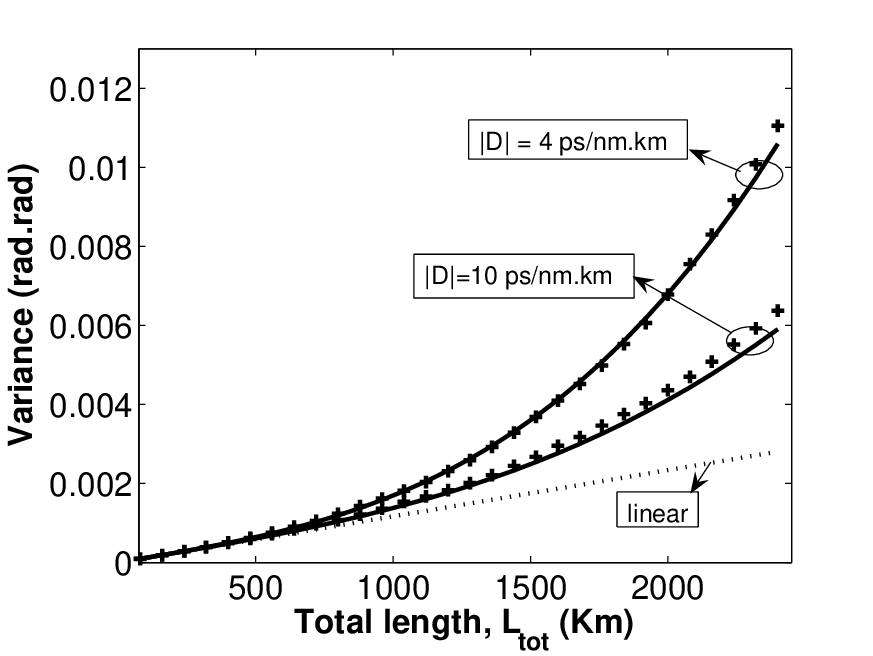}
\caption{Phase variance dependence on the total length of the transmission line. Peak launch power $=2\text{ mW}$. Solid line and + marks show the analytical and numerical simulation results, respectively. The dotted line shows the analytical results when fiber nonlinearity is absent, which is independent of dispersion. $\text{DOF}=14$ is used for analytical results.}
\label{fig:1}
\end{figure}

To estimate the number of DOFs required when a matched filter ($f_{0}=21.19\text{ GHz}$) is used, in Fig. 2, we have plotted the phase variance as a function of length of transmission line for various DOFs using (82). From Fig. 2, we see that the phase variance does not change as the number of DOFs is changed from 6 ($J=2$) to 14 ($J=6$). However, there is about 10\% change in variance as the number of DOFs is changed from 2 to 6 when $|D|=4\text{ ps/nm}\cdot\text{km}$ and $L_{tot}=2400\text{ km}$, and the corresponding change in variance when $|D|=10\text{ ps/nm}\cdot\text{km}$ is 6\%. In Fig. 3, a Gaussian filter with $f_{0}=42.38\text{ GHz}$ which has a bandwidth twice that of a matched filter is used at the receiver. In this case, we see that two DOFs are not sufficient to describe the impact of noise on the phase variance. The errors introduced by using 2, 6, and 10 DOFs are 30\%, 4\% and 1\%, respectively for $|D|=4\text{ ps/nm}\cdot\text{km}$ and $L_{tot}=2400\text{ km}$. As the filter bandwidth increases, higher order noise modes and noise fields due to nonlinear mixing of the signal and higher order noise modes occupy the pass band of the filter. Therefore, as the filter bandwidth increases, the variance of linear phase noise as well as nonlinear phase noise increases.

\begin{figure}[htbp]
\centering
\includegraphics[width=0.85\linewidth]{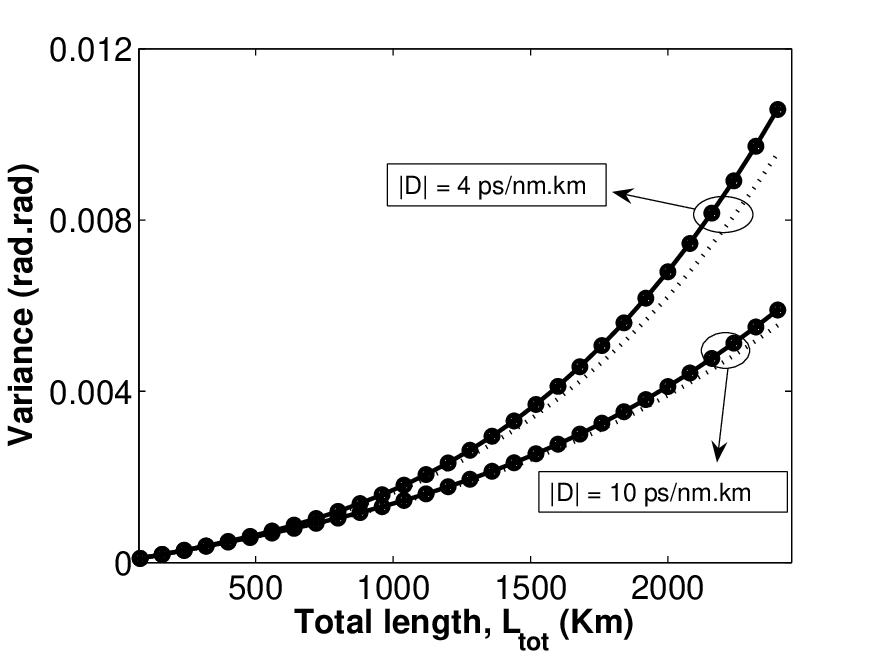}
\caption{Dependence of variance on the DOFs with a matched filter. Dotted line, circles, +, and solid line show the analytical results with DOF 2, 6, 10, and 14, respectively. Other parameters are same as that of Fig. 1.}
\label{fig:2}
\end{figure}

\begin{figure}[htbp]
\centering
\includegraphics[width=0.85\linewidth]{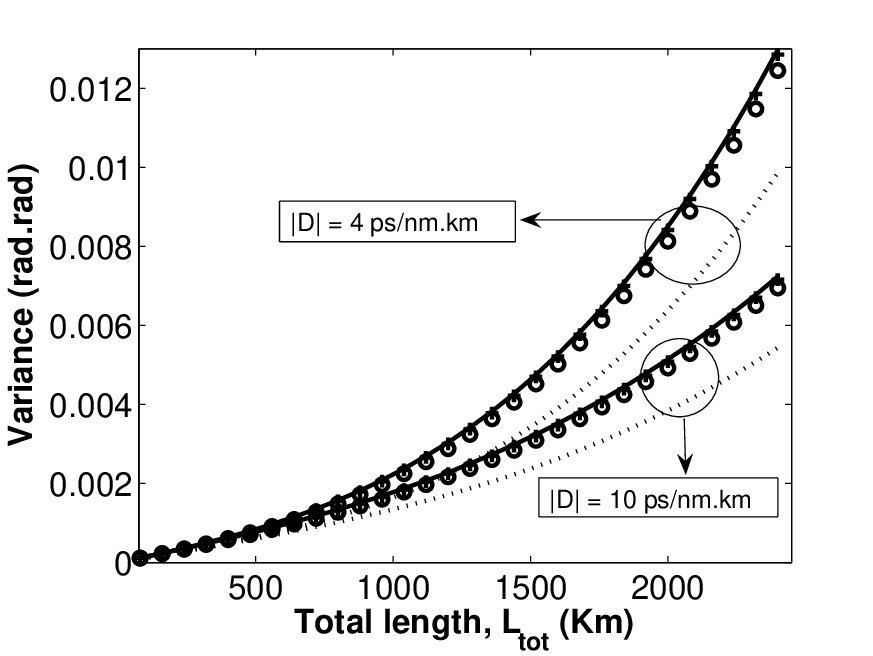}
\caption{Dependence of variance on the DOFs with a Gaussian filter with $f_{0}=42.38\text{ GHz}$. Dotted line, circles, +, and solid line show the analytical results with DOF 2, 6, 10, and 14, respectively. Other parameters are same as that of Fig. 1.}
\label{fig:3}
\end{figure}

Fig. 4 shows the dependence of phase variance on the launch power. When the launch power is low, the linear phase noise dominates (because of $1/E$ dependence in (82)). At high launch power, nonlinear phase noise becomes significant (because of $E$ dependence in (82)). The optimum launch power is calculated to be 1.8 mW using (40) which is in agreement with numerical simulations. At high launch powers ($> 4\text{ mW}$), there is a small discrepancy between the analytical results and simulation results which is due to the fact that we have ignored the terms containing $\gamma^{2}$ and higher. The first order perturbation theory is known to become inaccurate at large launch powers and/or longer transmission distance. It may be possible to increase the accuracy of the calculations using the multiple-scale approaches of [23], [24], and [30] when the dispersion map is periodic. Alternatively, a second-order perturbation theory [28] could be used which is shown to be quite accurate for the description of SPM and XPM for the range of launch powers and transmission distances of practical interest.

\begin{figure}[htbp]
\centering
\includegraphics[width=0.85\linewidth]{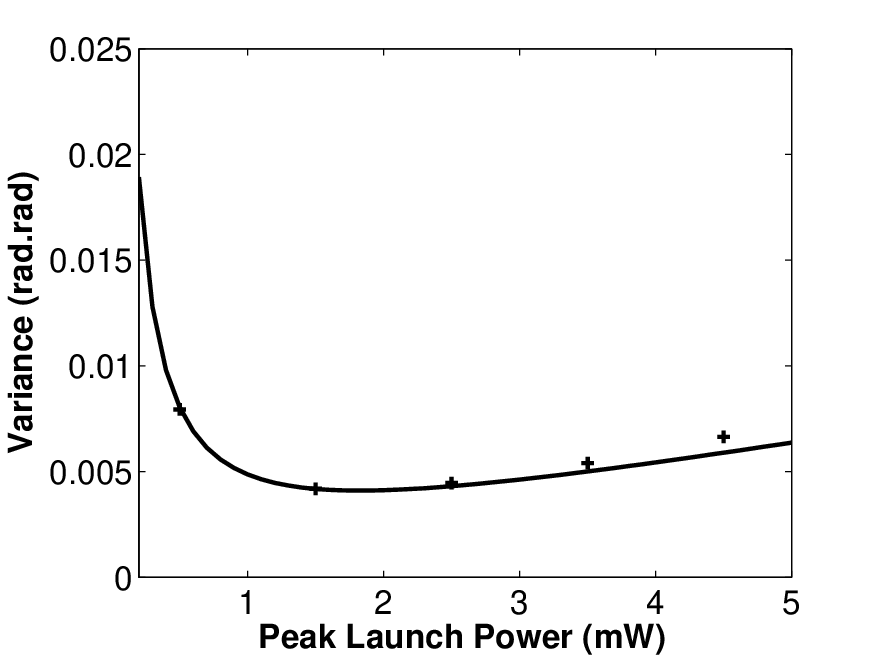}
\caption{Dependence of phase variance on peak launch power. Matched filter is used. Solid and + show the analytical and numerical simulation results, respectively. $L_{tot}=2400\text{ km}$ and $|D|=4\text{ ps/nm}\cdot\text{km}$. $\text{DOF}=14$ is used for analytical results.}
\label{fig:4}
\end{figure}

Next, we consider a dispersion map with two types of transmission fibers within an amplifier spacing. Let $D_{1}$ and $D_{2}$ be the dispersion parameters of these fibers and, $l_{1}$ and $l_{2}$ be their respective lengths. The average dispersion of these fibers is
\begin{equation}
D_{av}=\frac{(D_{1}l_{1}+D_{2}l_{2})}{(l_{1}+l_{2})}.
\end{equation}
The dispersion of the transmission fibers is compensated by pre- and postcompensating fibers. The dispersion coefficients and lengths of pre- and postcompensating fibers are so selected that the total accumulated dispersion before decision is zero. Fig. 5 shows the dependence of $g_{fr}(L_{1})$ on the average dispersion, $D_{av}$ for different values of $D_{1}$. $g_{fr}(L_{1})$ is the real part of $g_{f}(L_{1})$ defined in (27) which is proportional to the nonlinear phase acquired due to propagation from $L_{1}$ to $L_{tot}$. We have assumed the following parameters in Fig. 5. The dispersion parameter of the pre- and postcompensating fiber, $D_{pre}=D_{post}=-100\text{ ps/nm}\cdot\text{km}$, $l_{1}=l_{2}=40\text{ km}$, inline amplifier spacing $=l_{1}+l_{2}=80\text{ km}$, transmission distance (excluding lengths of pre- and postcompensation fibers), $L_{tr}=2400\text{ km}$ and launched peak power $=2\text{ mW}$. 50\% of the total accumulated dispersion of the transmission link is compensated using the precompensating fiber. As can be seen in Fig. 5, $g_{fr}$ decreases as $D_{av}$ or $|D_{1}|$ increases. Solid line in Fig. 6 shows the phase variance calculated from (77) and (34) and + shows the numerical simulation results. The phase variance also shows the similar dependence on $D_{av}$ and $D_{1}$ as $g_{fr}$. As $D_{av}$ and/or $|D_{1}|$ increases, the nonlinear contribution to the phase variance becomes quite small. However, in this case, pulses significantly broaden and overlap with neighboring pulses and it is likely that the ASE-induced nonlinear phase noise due to IXPM becomes important. This would be the subject of a future investigation.

\begin{figure}[htbp]
\centering
\includegraphics[width=0.85\linewidth]{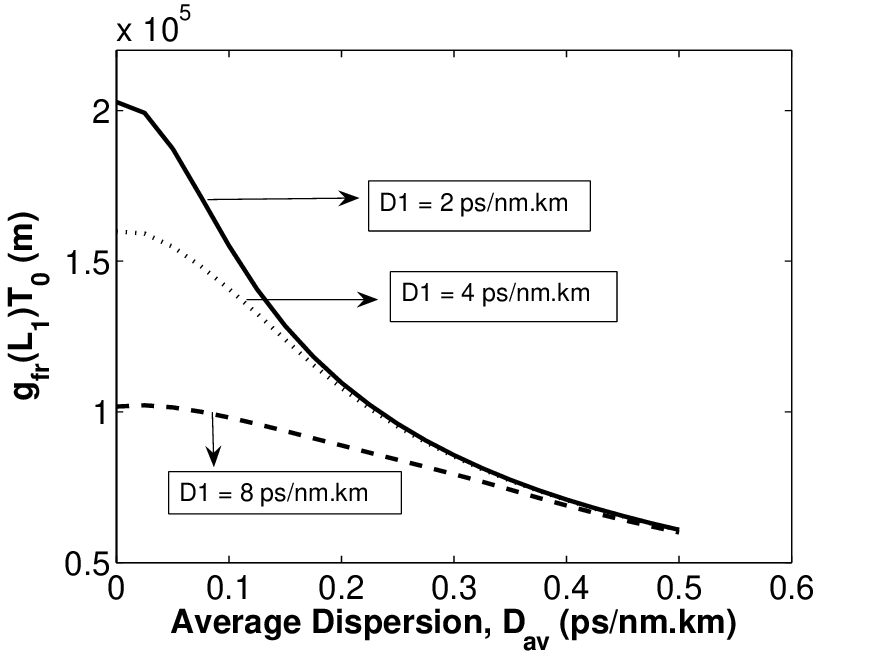}
\caption{Dependence of $g_{fr}(L_{1})$ on the average dispersion, $D_{av}$ and the local dispersion, $D_{1}$. Matched filter is used. Total transmission distance, $L_{tr}$ (excluding pre- and post-compensation fiber) $=2400\text{ km}$, peak power $=2\text{ mW}$, location of the first inline amplifier, $L_{1}=0.5 D_{av} L_{tr} / D_{pre}$ and $\text{DOF}=14$.}
\label{fig:5}
\end{figure}

\begin{figure}[htbp]
\centering
\includegraphics[width=0.85\linewidth]{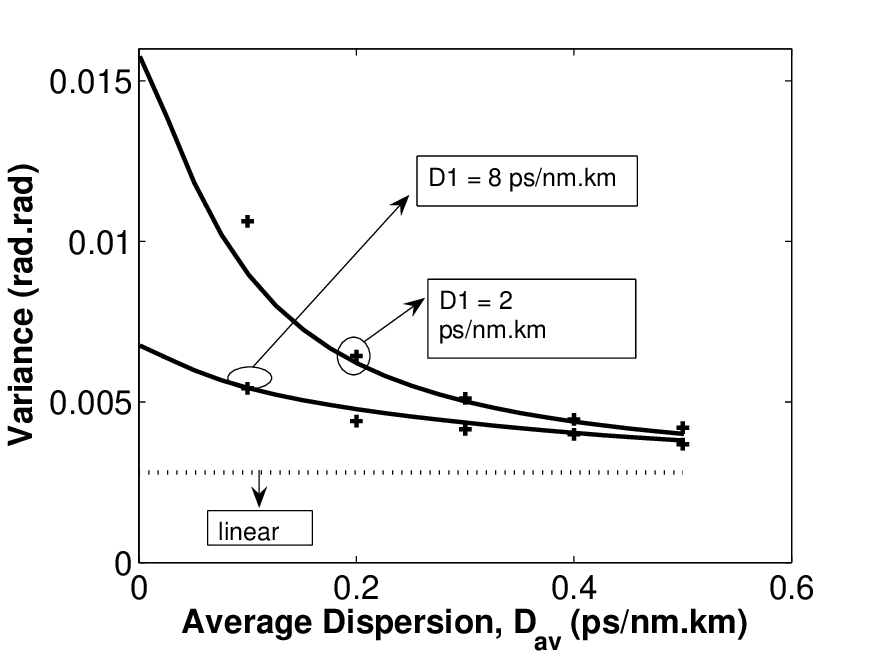}
\caption{Dependence of phase variance on the average dispersion, $D_{av}$ and the local dispersion $D_{1}$. Solid line and + show the analytical (with $J=6$) and numerical simulation results, respectively. Dotted line shows the analytical results for the case of $\gamma=0$ with $J=6.$ The other parameters are same as that of Fig. 5.}
\label{fig:6}
\end{figure}

\section{CONCLUSION}
We have developed analytical expressions for the linear and nonlinear phase variance due to SPM. The noise field is represented using Gauss-Hermite orthonormal set. In the first part, only two degrees of freedom (DOFs) of the noise field is considered which leads to simple analytical expressions for the phase variance. The in-phase component of the noise mode leads to energy fluctuations and consequent nonlinear phase fluctuations, and the quadrature component leads to linear phase noise consistent with [1]. In the second part, the results are extended to arbitrary DOFs and Gaussian filters with arbitrary bandwidths are inserted at the end of the transmission line. Nonlinear mixing of higher order noise modes with the signal pulse leads to new noise fields which enhance the nonlinear phase noise. Higher order noise modes propagate linearly and contribute to the linear phase noise. However, when a matched filter is used, higher order modes are orthogonal to the signal and linear phase noise is not enhanced. We have found that the maximum error introduced by including only two DOFs is about 10\% in our parameter space when a matched filter is used. However, this error could increase for longer transmission distance/higher launch power. When the filter bandwidth exceeds the signal bandwidth, higher order noise modes become more important which contribute to linear as well as nonlinear phase noise. We have also developed an analytic expression for the optimum launch power which is in good agreement with numerical simulations.

\appendices
\section{GAUSS-HERMITE SOLUTIONS}
Consider the Schr\"{o}dinger equation given by (1) with $\gamma=0$ and $R=0.$ Using the lens-like transformation [31]
\begin{equation}
u(z,t)=\sqrt{p(z)}\nu[p(z)t,z]\exp\left(iC(z)\frac{t^{2}}{2}\right)
\end{equation}
in (1) with $\gamma=R=0$ leads to [32]
\begin{multline}
i\left(\frac{\partial\nu}{\partial z}+K_{0}\tau\frac{\partial\nu}{\partial\tau}\right)-\frac{\beta_{2}(z)p^{2}(z)}{2}\frac{\partial^{2}\nu}{\partial\tau^{2}} \\
-\frac{(\dot{C}-C^{2}\beta_{2})\tau^{2}}{2p^{2}}\nu-\frac{iK_{0}}{2}\nu=0
\end{multline}
where $\tau=p(z)t$ and ``$\cdot$'' denotes the differentiation with respect to the argument. Dividing (88) by $\beta_{2}p^{2}$ we get
\begin{align}
K_{0} &= \frac{\dot{p}-C\beta_{2}p}{p} \\
i\frac{\partial\nu}{\partial z'}&-\frac{1}{2}\frac{\partial^{2}\nu}{\partial\tau^{2}}+\frac{K_{1}\tau^{2}\nu}{2}=0
\end{align}
where
\begin{align}
z' &= \int_{0}^{z}\beta_{2}(s)p^{2}(s)ds \\
K_{1} &= \frac{C^{2}\beta_{2}-\dot{C}}{\beta_{2}p^{4}}.
\end{align}

If we choose
\begin{equation}
K_{1}=\frac{C^{2}\beta_{2}-\dot{C}}{\beta_{2}p^{4}}=\text{constant}(=1)
\end{equation}
and
\begin{equation}
K_{0}=\frac{\dot{p}-C\beta_{2}p}{p}=0
\end{equation}
we obtain
\begin{equation}
i\frac{\partial\nu}{\partial z'}-\frac{1}{2}\frac{\partial^{2}\nu}{\partial\tau^{2}}+\frac{\tau^{2}\nu}{2}=0.
\end{equation}

Let
\begin{equation}
\nu(z',\tau)=\phi_{j}(\tau)\exp(i\lambda_{j}z'), \quad j=0,1,2,\dots
\end{equation}
Substituting (96) in (95), we obtain
\begin{equation}
\frac{1}{2}\frac{d^{2}\phi_{j}}{d\tau^{2}}-\frac{\tau^{2}\phi_{j}}{2}=-\lambda_{j}\phi_{j}.
\end{equation}

The above equation is the eigenvalue equation corresponding to the quantum harmonic oscillator. The eigenfunction $\phi_{j}$ equals Gauss-Hermite functions of (41) and the eigenvalue $\lambda_{j}$ is given by
\begin{equation}
\lambda_{j}=2j+1.
\end{equation}

The constraints imposed in (93) and (94) on $p(z)$ and $C(z)$ with the initial condition $p(0)=1/T_{0}$ and $C(0)=0$ lead to the following solution:
\begin{equation}
p(z)=\frac{T_{0}}{\sqrt{T_{0}^{4}+S^{2}(z)}}, \quad C(z)=\frac{S(z)p^{2}(z)}{T_{0}^{2}}.
\end{equation}
Note that (99) is same as (9) which is obtained using Fourier transform technique. Substituting (99) in (91), we find
\begin{equation}
z'=\frac{1}{2}\tan^{-1}\left[\frac{S(z)}{T_{0}^{2}}\right].
\end{equation}

Combining (87), (96) and (98), we obtain the final solution
\begin{equation}
u(z,t)=\sqrt{p(z)}k_{j}H_{j}(pt)\exp\left[iC(z)\frac{t^{2}}{2}+i(2j+1)\theta_{0}(z)\right].
\end{equation}
To our knowledge, the above solution was first obtained by Lazaridis et al. [22] for temporal pulse propagation using a different approach.

\section{SOLUTION OF (58)}
Consider the terms on the right hand side of (58). They can be written as
\begin{align}
T_{1}(z,t) &= |\psi_{0}|^{2}\psi_{j} \nonumber \\
&= \frac{p^{3/2}k_{j}}{\sqrt{\pi}}\exp\left[-\frac{(p_{1}t)^{2}}{2}+i\theta_{j}(z)\right]\frac{d^{j}[\exp(-\tau^{2})]}{d\tau^{j}}
\end{align}
where
\begin{align}
p_{1}^{2} &= p^{2}-iC, \quad \theta_{j}(z)=(2j+1)\theta_{0}(z), \quad \tau=pt \\
T_{2}(z,t) &= \psi_{0}^{2}\psi_{j}^{*} \nonumber \\
&= \frac{p^{3/2}k_{j}}{\sqrt{\pi}}\exp\left\{-\frac{(p_{1}t)^{2}}{2}+i[2\theta_{0}(z)-\theta_{j}(z)]\right\} \nonumber \\
&\quad \times \frac{d^{j}[\exp(-\tau^{2})]}{d\tau^{j}}.
\end{align}

We take the Fourier transform of (58) to obtain
\begin{multline}
i\frac{d\tilde{\psi}_{j}^{(1)}}{dz}+\frac{\beta_{2}(z)(2\pi f)^{2}}{2}\tilde{\psi}_{j}^{(1)}=-E\exp[-w(z)] \\
\times[2n_{j}\tilde{T}_{1}(z,f)+n_{j}^{*}\tilde{T}_{2}(z,f)]
\end{multline}
where
\begin{align}
\tilde{T}_{l}(z,f) &= \mathcal{F}[T_{l}(z,t)], \quad l=1,2 \\
\tilde{\psi}_{j}^{(1)}(z,f) &= \mathcal{F}[\psi_{j}^{(1)}(z,t)].
\end{align}

Noting that
\begin{align}
\mathcal{F}\left[\frac{d^{j}(\exp(-\tau^{2}))}{d\tau^{j}}\right] &= \int \frac{d^{j}[\exp(-\tau^{2})]}{d\tau^{j}}\exp(i2\pi ft)dt \nonumber \\
&= \frac{\sqrt{\pi}(-i2\pi f)^{j}\exp(-\frac{f^{2}\pi^{2}}{p^{2}})}{p^{j+1}}
\end{align}
$\tilde{T}_{1}(z,f)$ can be simplified as
\begin{multline}
\tilde{T}_{1}(z,f)=\frac{k_{j}(-i2\pi)^{j}p_{2}\sqrt{2}\exp\left(-\frac{\pi^{2}f^{2}}{p_{3}^{2}}\right)+i\theta_{j}}{\sqrt{\pi}p_{1}p^{j-1/2}} \\
\times\sum_{k=0}^{j}\frac{(2f)^{k}p_{2}^{j+k}I(j-k)j!}{p_{1}^{2k}\pi^{j-k}k!(j-k)!}
\end{multline}
where
\begin{align}
I(j) &= \int f^{j}\exp(-f^{2})df \nonumber \\
&= \frac{(j-1)!\sqrt{\pi}}{2^{j-1}(\frac{j}{2}-1)!}, \quad \text{if } j \text{ is even, } \neq 0 \\
&= \sqrt{\pi}, \quad \text{if } j=0 \\
&= 0, \quad \text{if } j \text{ is odd}
\end{align}
and
\begin{equation}
p_{2}^{2}=\frac{p^{2}(p^{2}-iC)}{3p^{2}-iC}, \quad p_{3}^{2}=\frac{3p^{2}-iC}{2}.
\end{equation}

Considering only the term containing $\tilde{T}_{1}(z,f)$ in (105), it can be solved to obtain
\begin{multline}
\tilde{\psi}_{j1}^{(1)}(L_{tot},f)=2in_{j}E\int_{L_{m}}^{L_{tot}}\exp[-w(z)-i2S(z)(\pi f)^{2}] \\
\times\tilde{T}_{1}(z,f)dz.
\end{multline}

Multiplying $\tilde{\psi}_{j1}^{(1)}(L_{tot},f)$ by the filter transfer function $H_{G}(f)$ ((59)) and inverse Fourier transforming, we obtain the sample at $t=0$ as
\begin{equation}
\psi_{j1}^{(1)}(L_{tot},0)=n_{j}EA_{j}
\end{equation}
where
\begin{align}
A_{j} &= \int_{L_{m}}^{L_{tot}}\eta_{j}(z)dz \\
\eta_{j}(z) &= \sum_{k=0}^{j}\frac{\Delta_{j}(z)}{\beta^{(k+1)/2}}\times\frac{2^{k}p_{2}^{j+k}j!I(k)I(j-k)}{p_{1}^{2k}\pi^{j-k}k!(j-k)!} \\
\Delta_{j}(z) &= 2i\exp[-w(z)-\beta(z)+i\theta_{j}]\sqrt{\frac{2p}{\pi}}\frac{k_{j}p_{2}(-i2\pi)^{j}}{p_{1}p^{j}} \\
\beta(z) &= 2\pi^{2}\left[\frac{1}{3p^{2}(z)-iC(z)}+iS(z)\right]+\frac{1}{2f_{0}^{2}}.
\end{align}

Next, let us consider the second term on the right hand side of (105). Comparing (102) and (104), we see that they are the same except for the phase factor, i.e.,
\begin{equation}
T_{2}(z,t)=T_{1}(z,t)\exp\{2i[\theta_{0}(z)-\theta_{j}(z)]\}.
\end{equation}
Therefore, the solution of (105) considering only the second term on the right hand side is
\begin{multline}
\tilde{\psi}_{j2}^{(1)}(L_{tot},f)=in_{j}^{*}E\int_{L_{m}}^{L_{tot}}\exp[-w(z)-i2S(z)(\pi f)^{2}] \\
\times \exp\{2i[\theta_{0}(z)-\theta_{j}(z)]\}\tilde{T}_{1}(z,f)dz.
\end{multline}

Proceeding as before, the sample at $t=0$ of the filter output is
\begin{equation}
\psi_{j2}^{(1)}(L_{tot},0)=n_{j}^{*}EB_{j}
\end{equation}
where
\begin{align}
B_{j} &= \int_{L_{m}}^{L_{tot}}\xi_{j}(z)dz \\
\xi_{j}(z) &= \sum_{k=0}^{J}\frac{\Delta_{j}'(z)}{\beta^{(k+1)/2}}\times\frac{2^{k}p_{2}^{j+k}j!I(k)I(j-k)}{p_{1}^{2k}\pi^{j-k}k!(j-k)!} \\
\Delta_{j}'(z) &= \Delta_{j}\frac{\exp\{2i[\theta_{0}(z)-\theta_{j}(z)]\}}{2}.
\end{align}

The complete solution is
\begin{align}
\psi_{jf}^{(1)} &= \psi_{j1}^{(1)}(L_{tot},0)+\psi_{j2}^{(1)}(L_{tot},0) \nonumber \\
&= E(n_{j}A_{j}+n_{j}^{*}B_{j}).
\end{align}

\begin{IEEEbiography}{Shiva Kumar}
 received the B.E. degree in electronics and communications from Mysore University, India, in 1988, the M.S. degree in 1990, the Ph.D. degree in 1994 in electrical communication engineering from the Indian Institute of Science, Bangalore, and the Ph.D. degree in 1997 in communications engineering from Osaka University, Osaka, Japan.

He was a Postdoctoral Fellow at the University of Jena, Germany, supported by the Alexander von Humboldt Foundation from 1997--1998. He was with Corning, Inc., New York, as a Senior Research Scientist (1998--2002) and as a Research Associate (2002--2003). He joined McMaster University, Hamilton, ON, Canada, in 2003. Currently, he is an Associate Professor in the Department of Electrical and Computer Engineering, McMaster University. His current research interests include optical communication, solitons, nonlinear optics, and photonic devices.
\end{IEEEbiography}

\end{document}